\documentclass[a4paper]{article}
\usepackage[margin=1cm]{caption}
\usepackage[margin=3.5cm]{geometry}
\usepackage{graphicx}
\usepackage{float}
\usepackage[hidelinks]{hyperref}
\usepackage[frozencache,cachedir=.]{minted}
\usepackage[natbibapa]{apacite}
\usepackage{inconsolata}
\usepackage{xcolor}
\usepackage{censor}
\usepackage{setspace}

\definecolor{bg}{rgb}{0.97,0.97,0.97}

\setminted{
    linenos,
    fontsize=\small,
    bgcolor=bg,
}

\renewcommand{\censor}[1]{#1}

\title{\textbf{Tidy simulation}\\Designing robust, reproducible, and scalable Monte Carlo simulations}
\author{\censor{Erik-Jan van Kesteren\\\emph{Utrecht University, Department of Methodology \& Statistics}}}
\date{}

\begin{document}

\maketitle
\begin{abstract}
    Monte Carlo simulation studies are at the core of the modern applied, computational, and theoretical statistical literature. Simulation is a broadly applicable research tool, used to collect data on the relative performance of methods or data analysis approaches under a well-defined data-generating process. However, extant literature focuses largely on design aspects of simulation, rather than implementation strategies aligned with the current state of (statistical) programming languages, portable data formats, and multi-node cluster computing.
    
    In this work, I propose tidy simulation: a simple, language-agnostic, yet flexible functional framework for designing, writing, and running simulation studies. It has four components: a tidy simulation grid, a data generation function, an analysis function, and a results table. Using this structure, even the smallest simulations can be written in a consistent, modular way, yet they can be readily scaled to thousands of nodes in a computer cluster should the need arise. Tidy simulation also supports the iterative, sometimes exploratory nature of simulation-based experiments. By adopting the tidy simulation approach, researchers can implement their simulations in a robust, reproducible, and scalable way, which contributes to high-quality statistical science.
\end{abstract}

\section{Introduction}
Since the advent of statistical computing, Monte Carlo simulations have become one of the pillars of modern statistical research. Simulations are ``in silico'' experiments, one of the main data collection tools of the methodologist. Accordingly, simulations are widely used for various goals, for example to support the research and development of new statistical methods \citep{tibshirani1996regression}, to illustrate a theoretical point within a broader argument \citep{box1976science}, to benchmark when and where an existing statistical method works better than another \citep{mackinnon1995simulation}, for ``a priori'' power analysis for complex models where analytical power is infeasible to compute \citep{lakens2021simulation, constantin2023general}, to dispel common myths in practical data science questions \citep{carriero2025harms}, or to validate custom computational models in Bayesian modeling workflows \citep{gelman2020bayesian}.

In this development of simulation as a core research tool in the field of statistics, there have been several excellent publications which detail how to create simulations for one or more of these goals \citep{morris2019using, burton2006design, boulesteix2020introduction}. These focus on how to design, set up, and run a simulation study, and they provide a stepping stone for researchers to produce reliable simulations, as well as for practitioners to understand their conclusions.

Yet, despite these efforts, there is a wide range of approaches of how researchers implement and scale up statistical simulations in practice. To a large extent, this is unavoidable, as simulations are used as a tool to create new knowledge -- thus, each simulation has its own computational and practical idiosyncrasies. Nevertheless, for researchers it is inefficient to rediscover data structures, processing pipelines, and project structures that facilitate reproducibility, transparency, scalability, and robustness in the modern computational statistics world. Apart from a few notable exceptions \citep{lee2019using, hussey2025tidy}, there is currently little guidance on these implementation choices.

Additionally, extant literature lacks focus on practical aspects such as debugging and error handling. It also disregards the iterative process of running simulation studies: often, analyzing (preliminary) simulation results is a process of exploratory data analysis \citep{tukey1977exploratory}. Notably, in simulations it is relatively easy to collect new ``data'' as part of this process of discovery. As a result, in creating and running a simulation study, the researcher often wants to add more factors to vary in the data-generating process, or more levels to a certain factor. Having a consistent, structured simulation design can help support this iterative process, making it easy to add new components or remove unnecessary ones. As with any experiment, this should be done transparently and considering carefully the pitfalls of overfitting.

\vspace{12pt}

The present paper proposes \textit{tidy simulation}: a framework for designing simulations using principles from tidy data \citep{wickham2014tidy} and developments in programming languages, multi-core and distributed computing, and modern cross-language data formats \citep[e.g.,][]{apacheparquet}. Broadly, this proposal provides a modern answer to the question of how to implement a Monte Carlo simulation for statistical or methodological research and practice. Open example code implementing this framework in Python and R is available in the online \href{https://github.com/vankesteren/tidy_simulation}{\texttt{code repository}}. For all its benefits, simulation should be a commonplace methodology with a broad user-base, not a niche skill reserved only for the most technologically adept statisticians. Using the tidy simulation framework, simulation can become a more standardized, portable research tool for answering methodological questions.

\begin{figure}[h]
    \centering
    \includegraphics[width=0.8\linewidth]{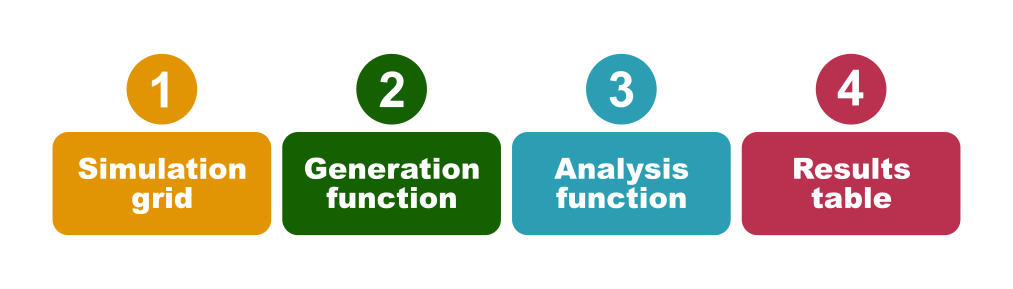}
    \caption{Four sequential components of a tidy simulation. The simulation grid is a tidy data frame with the settings for each iteration of the simulation, the generation function creates simulated data through random sampling, the analysis function applies the methods of interest to this data, and the results table compiles the outputs from the analysis, again in a tidy data format.}
    \label{fig:tidy}
\end{figure}

\section{Statistical simulations}

Statistical simulations are experiments, designed to gather evidence towards answering a methodological research question. The evidence in question compares the performance of different estimands or methods of interest, for example  whether a novel method is better than an existing approach. Statistical simulations are special in that the data used is not collected from real-world measurements, but rather simulated, or sampled from a well-defined, known stochastic data-generating process.

Following the methodology adopted by \citet{morris2019using} and \citet{boulesteix2020introduction}, the key elements determining the design of a statistical simulation are as follows:

\begin{itemize}
    \item (Aims) What is the research question?
    \item (Data-generating mechanism) How is the stochastic data generated?
    \item (Estimand \& Method) Which method do we apply to the generated data?
    \item (Performance measures) Which metric are we interested in comparing and presenting? 
\end{itemize}

This list is referred to with the acronym ADEMP. In practice, these design choices may need be considerably edited during the implementation phase in an iterative, exploratory process -- perhaps with the exception of the main aim. For example, it may turn out that a certain model is incompatible with a particular data-generating mechanism, or that an additional model should be added to the comparison. Thus, while it is beneficial to have a good understanding of each of these key components before writing a simulation, its implementation needs to be modular such that changes are readily implemented.

Additionally, like in real-world experiments, statistical simulations contain different generation or analysis conditions -- sometimes referred to as ``factors to vary''. While it is not immediately obvious from the ADEMP checklist, these factors to vary are among the most important elements to present to the reader of a study. Common factors are the sample size or the number of variables in the data-generating process, or the hyperparameters of a statistical analysis method (such as the number of latent classes or the optimizer). It is common to report these factors in a table, which allows readers to quickly view the simulation settings.

For a more in-depth overview on the different design choices of a simulation study, I refer the reader to the comprehensive work by \citet{morris2019using}. Beyond introducing the ADEMP checklist, it provides further context and detailed suggestions for each design choice, such as which performance metric to choose for various data settings, and how to ensure the different methods are compared in a fair way. The present work on tidy simulation is complementary to earlier reviews, rather focusing on robust, reproducible, and scalable implementation.

\section{Tidy simulation}

A tidy simulation is a statistical experiment using the widely used concept of \emph{tidy data} as its main data structure. In tidy data, (a) each variable is a column, each column is a variable; (b) each observation is a row, each row is an observation; and (c) each value is a cell, each cell is a single value \citep{wickham2014tidy}. Tidy simulations define a tidy grid of conditions and a results table computed via modular functions. This happens through four sequential components (Figure \ref{fig:tidy}): 

\begin{enumerate}
    \item \textbf{Simulation grid}: a tidy data frame indicating the settings for each iteration of a simulation.
    \item \textbf{Generation function}: takes in parameters from the simulation grid and outputs a simulated dataset -- ideally in a tidy format.
    \item \textbf{Analysis function}: accepts a single dataset and additional parameters from the grid, runs a method on this dataset, and yields one or more outputs of interest.
    \item \textbf{Results table}: a tidy data frame with one row per simulation setting in the simulation grid, and one or more columns containing the outcomes of interest.
\end{enumerate}

\begin{figure}[H]
    \centering
    \includegraphics[width=\linewidth]{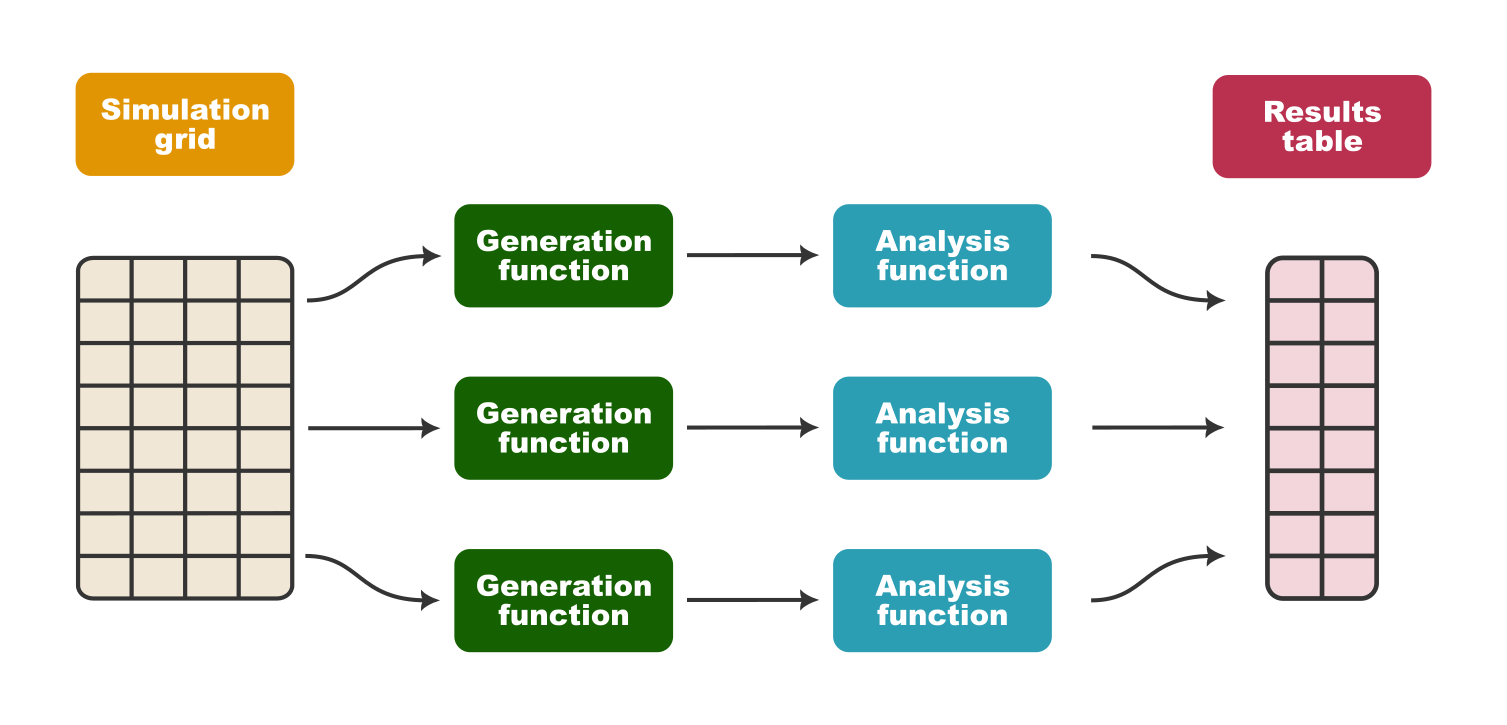}
    \caption{Tidy simulations are embarrassingly parallel, as each row in the simulation grid and results table represents a single simulation setting (tidy data), meaning it is independent of the others.}
    \label{fig:scale}
\end{figure}

This sequential, functional approach makes tidy simulations \textbf{modular}: edits can be made in one part, while the other components remain the same. In addition, each row in the simulation grid independently leads to a row in the results table through data generation and the analysis function. This independence also makes tidy simulations \textbf{scalable}: each row or chunk of rows can be readily run in parallel, in a separate process (Figure \ref{fig:scale}), or even on separate nodes of a high-performance computer cluster running standard scheduling software \citep[e.g., SLURM;][]{yoo2003slurm}.

Several opinionated software frameworks currently exist for performing simulations based on tidy principles, specifically in the R community \citep[e.g., \texttt{simpr}, \texttt{simChef}, or \texttt{simhelpers};][]{brown2022simpr, duncan2024simchef, joshi2025simhelpers}. However, like \citep{hussey2025tidy}, the present work proposes a general framework rather than a particular implementation. This adds flexibility and cross-language compatibility: defining and decoupling each component rather than creating a single unified framework allows for solving idiosyncratic problems such as memory limits, incremental running, and scaling to large clusters.

The following subsections describe in detail what each of the tidy simulation components entail, as well as practical considerations around implementing these components. To do so, we follow a running example of power analysis for pre-post treatment analysis -- a contentious topic, with a long statistical history \citep[see, e.g.,][and many more]{huck1975using, norman1989issues, jamieson1995measurement, liu2009should, kisbu2013monte}.

\begin{quote}
    \textbf{Running example}. A researcher wants to analyze the treatment effect in a randomized controlled trial with a binary treatment and two measurement occasions: pre-treatment and post-treatment. Data collection is very expensive, so they want to know what power they have to detect the treatment effect with few samples and different linear models: either change score (post-pre) as an outcome or the post-treatment measurement only as an outcome; either corrected for the pre-treatment values or not.
\end{quote}

\subsection{Simulation grid}
\label{sec:simgrid}
The simulation grid is a tidy table that contains all the combinations of the factors to vary -- henceforth called \emph{simulation factors} -- repeated for each iteration. These commonly contain discrete values (such as sample size or the number of variables), continuous values (such as standardized effect size or the proportion of variance explained), and categorical values (such as distribution type or analysis method), but are not restricted to those. For our running example, the simulation factors are displayed in Table \ref{tab:factors}, with an indication of their values.

\begin{table}[H]
    \centering
    \begin{tabular}{ll} 
        \hline
        \textbf{Simulation factor} & \textbf{Levels / values} \\
        \hline
        Sample size & \texttt{4, 5, 6, ..., 19} \\
        Treatment effect size & \texttt{0, 0.1, 0.2, ..., 1.0} \\
        Outcome type & \texttt{Post-treatment, change score} \\
        Adjustment & \texttt{Uncorrected, baseline-corrected} \\
        \hline
    \end{tabular}
    \caption{Simulation factors in the running example simulation.}
    \label{tab:factors}
\end{table}

While not explicitly stated in Table \ref{tab:factors}, there are two distinct types of simulation factors. The sample size and the treatment effect size are \emph{data generation factors} that alter the data-generating process; the outcome type and correction are \emph{analysis factors} that adjust the methods to be evaluated. This ties in closely with the components described in sections \ref{sec:datgen} and \ref{sec:analfun}, respectively. 

\subsubsection{Grid creation}
The simulation grid is produced by creating a table representing a full factorial design of the simulation factors: each possible combination of their levels appears in a structured order. There are several functions in different programming languages to support creating such a design: R defines \texttt{expand.grid()} or \texttt{expand\_grid()} in tidyr; in Python there is \texttt{itertools.product()}, \texttt{meshgrid()} from numpy \citep{harris2020array}, or \texttt{expand\_grid()} from the package polarsgrid \citep{vankesteren2025polarsgrid}; MATLAB has \texttt{ndgrid()}; and Julia has \texttt{Iterators.product()}. Alternatively, this can be achieved using SQL-style cross-joins (e.g., in Stata, SAS, or databases), or nested loops in languages that do not have convenience functions for this (e.g., Fortran). 

\begin{minted}{python}
from polarsgrid import expand_grid
simulation_grid = expand_grid(
    sample_size=[4, 5, 6, 7, 8, 9, 10, 11, 12, 13, 14, 15, 16, 17, 18, 19],
    effect_size=[0.0, 0.1, 0.2, 0.3, 0.4, 0.5, 0.6, 0.7, 0.8, 0.9, 1.0],
    outcome=["post", "change"],
    correction=[False, True],
    iteration=list(range(500)),
)
\end{minted}

Monte Carlo simulations require repeated sampling from the data-generating process -- commonly, 500 or 1000 repetitions. Since each row in the simulation grid should represent a single simulation according to the tidy data principles, the full factorial design should be stacked $N$ times, where $N$ is the number of iterations. The iteration number should also be included in creating the simulation grid, which enables partial results computation: a methodologist can start with computing 5 iterations of the full design by filtering the grid on the iteration number column, or by taking the first $1\%$ of rows. This allows preliminary analyses to be done on a subset of the results table (see Sections \ref{sec:restab} and \ref{sec:analysis}), supporting exploratory data analysis, with additional precision added later upon running and storing the remaining iterations.

Table \ref{tab:sim-grid} shows the simulation grid for the running example with 500 repetitions. Note that there are 352000 individual simulations even for this small simulation with only four factors. The size of the grid can explode with more complex simulations. To avoid memory problems, the data structure should be efficient: each factor should be encoded as a categorical data type (factor, categorical, or enum) and not as a string or character. If this is done and the grid is still too large --- many millions of rows are not uncommon in larger simulations --- the grid can also be stored on disk in a portable format of which individual rows can be read efficiently. A particularly efficient binary format for this is Apache parquet \citep{apacheparquet}, which is interoperable with many programming languages. Another option is to use database software with tight programming language integration; a modern portable option is duckdb \citep{raasveldt2019duckdb}. An advanced alternative to storing the grid on disk is to create a ``lazy'' table in which rows are only instantiated as needed. This is possible in the Polars dataframe library \citep{vink2025polars} using polarsgrid \citep{vankesteren2025polarsgrid}.

\begin{table}[ht]
    \centering
    \begin{tabular}{|c|cc|cc|c|}
         \hline
         &  \multicolumn{2}{c|}{\textit{Generation factors}}  & \multicolumn{2}{c|}{\textit{Analysis factors}}  & \\
         Row ID & Sample size  & Effect size & Outcome & Correction & Iteration \\
         \hline
         1 & 4 & 0.0 & post & false & 1 \\
         2 & 5 & 0.0 & post & false & 1 \\
         3 & 6 & 0.0 & post & false & 1 \\
         $\vdots$ & $\vdots$ & $\vdots$ & $\vdots$ & $\vdots$ & \\
         17600 & 19 & 1.0 & change & true & 250 \\
         17601 & 4 & 0.0 & post & false & 251 \\
         17602 & 5 & 0.0 & post & false & 251 \\
         $\vdots$ & $\vdots$ & $\vdots$ & $\vdots$ & $\vdots$ & \\
         351998 & 17 & 1.0 & change & true & 500 \\
         351999 & 18 & 1.0 & change & true & 500 \\
         352000 & 19 & 1.0 & change & true & 500 \\
         \hline
    \end{tabular}
    \caption{Example simulation grid for a statistical simulation with 500 iterations. The figure shows the first, middle, and last three rows of the long simulation grid.}
    \label{tab:sim-grid}
\end{table}

The example simulation study is small and can be efficiently run, but there are cases in which data generation is a computational bottleneck. For example, in agent-based modeling, a single dataset can take several seconds, minutes, or even hours to generate. In this case, it makes sense to retain only the \textit{data generation factors} in the simulation grid and to relegate the \textit{analysis factors} to the results table directly. This ensures that each analysis type is done on each generated dataset, and yields a ``wide'' analysis table, which can be readily transformed back to the ``long'' table shown here by pivoting the data frame. From an experimental design standpoint, this results in a repeated-measures design of the statistical experiment, as opposed to the independent-measures design of the example simulation. See Sections \ref{sec:restab} and \ref{sec:analysis} for a further description of this modification of tidy simulation.

\subsubsection{Grid post-processing}
After instantiating a full factorial simulation grid, some post-processing steps may be taken to further prepare the statistical simulation. Post-processing should not interfere with the tidy nature of the simulation grid.

One common operation is to remove or filter specific rows. This may be done, for example, when the experiment is very time-intensive -- in this situation a fractional factorial design \citep{gunst2009fractional} may provide enough information to support the research question. Another reason could be that certain combinations of parameter settings are inadmissible, such as when the number of groups is larger than the number of participants in a multilevel data-generating processes. 

The grid contains mainly (categorical) factors to vary, but additional metadata can be added as well. For example, it might be prudent for the purposes of reproducibility and debugging to add a data generation seed per row. This enables the methodologist to manually ``replay'' a row of an analysis exactly, and determine the source of an error in the data-generation function or non-convergence in the analysis function. Another potential inclusion may be references to external resources, which can be added through file paths; these are then loaded by either the data generation function or the analysis function.

\subsection{Data generation function}
\label{sec:datgen}

The data generation function is an explicit representation of the theoretical data-generating process in the Monte Carlo simulation. This function can take a wide range of forms, depending on the research question or the process under investigation. Generally, the data generation function takes in arguments from the simulation grid, such as the sample size or the effect size, and produces a single tidy dataset.

Since the data generation function will potentially be run millions of times, it needs to be efficient. Where possible, random sampling should be done with default or industry-standard implementations of probability distributions. For example, in R these are mostly built-in (e.g. \texttt{rnorm()}), Python has \texttt{numpy.random} and \texttt{scipy.stats} \citep{virtanen2020scipy} which define various distributions, and Julia has the excellent \texttt{Distributions.jl} package \citep{besancon2021distributions}. The random seed value from the simulation grid, if present, should also be properly handled. A simplified example in Python could be:

\begin{minted}{python}
import numpy as np
import polars as pl

def generate_data(sample_size: int, effect_size: float, seed: int):
    np.random.seed(seed)

    # sample pre and post variables
    treated = np.arange(sample_size) < (sample_size // 2)
    pre = np.random.normal(0, 3, sample_size)
    post = pre + np.random.normal(0, 0.3, sample_size)
    post[treated] += effect_size

    # return a tidy dataframe
    return pl.DataFrame({
        "id": np.arange(sample_size), 
        "treated": treated, 
        "pre": pre, 
        "post": post,
    })
\end{minted}

As the generation function grows in size, the need might arise to further modularize this function. For example, if there are three qualitatively different data generating processes under investigation, a methodologist might create three separate sub-functions and then call those from within the main data generation function:

\begin{minted}{python}
def generate_data(process: str, sample_size: int, effect_size: float, seed: int):
    if process == "a":
        return generate_a(sample_size, effect_size, seed)
    if process == "b":
        return generate_b(sample_size, effect_size, seed)
    if process == "c":
        return generate_c(sample_size, effect_size, seed)
\end{minted}

In addition to being efficient, the data generation function needs to be robust: it should be able to handle any combination of arguments present in the simulation grid, and where necessary it should handle errors gracefully. This can be done by raising appropriate, descriptive and clear errors, for which most programming languages have well-defined functionality. For example, in multilevel simulations (e.g., students within classrooms) there cannot be more groups than the total sample size, and this can be checked ahead of time:

\begin{minted}{python}
def generate_data(num_groups: int, sample_size: int):
    if num_groups >= sample_size:
        raise ValueError("Sample size should be larger than groups.")
    ...
\end{minted}

For some simulations it may not suffice to output only a tidy dataframe from the data generation function; generated "true" parameter values might also need to be passed on to the analysis function (section \ref{sec:analfun}) in order to compute bias or mean squared error. The tidy simulation framework is flexible enough that it allows for this: in Python or Julia, the function can return a tuple; in R, the function can return a list. Simplicity and legibility are valuable, so a generation function for linear regression data might also look like this:

\begin{minted}{python}
def generate_data(N: int, P: int, seed: int):
    np.random.seed(seed)
    X = np.random.standard_normal((N, P))
    b = np.random.normal(1, 1, P)
    y = X @ b + np.random.normal(0, 1, N)
    return X, y, b
\end{minted}

The outputs of the generation function are the inputs of the analysis function in the following section. These outputs thus define an \emph{interface} between these functions. Its stability is important for the modularity of the tidy simulation: if the interface remains the same (e.g., a data frame with a certain structure), the data-generating process can be updated without changing the analysis function and vice versa.

\subsection{Analysis function}
\label{sec:analfun}

The analysis function takes in a generated dataset (and any other output from the data generation function) along with arguments from the simulation grid, and produces the metrics or outcomes of interest which constitute a single row of the results table (Section \ref{sec:restab}). For a power analysis, this might be a \emph{p}-value; for a method comparison this might be relative bias \citep[refer to][for guidance on choosing the outcome]{morris2019using}. 

As with the generation function (Section \ref{sec:datgen}), the analysis function needs to be performant, as it will run many times. However, it needs even more attention towards robustness, as errors in statistical models might be probabilistic and data-dependent. For example, multicollinearity will lead to failure of ordinary least squares estimates of linear models \citep{alin2010multicollinearity}, full class separation in generated data with a binary outcome may lead to divergent parameter estimates in logistic regression \citep{heinze2006comparative}, particular edge-cases may arise in which mixed models do not converge \citep{eager2017mixed}, and certain datasets lead to inadmissible solutions in structural equation modeling \citep{paxton2001monte}. It is in the probabilistic nature of Monte Carlo simulations that even if these edge-cases are very unlikely to happen in any single generated dataset, they are highly likely to happen in the simulation as a whole. For example, if 99.999\% of cases from the data-generating process are not problematic, when data is generated 352000 times there is a probability of $1 - 0.99999^{352000} = 97\%$ of encountering at least one of these problems.

These data-dependent issues are particularly difficult to solve and debug, as they may occur at random. A few precautions can be taken to mitigate this issue. First and foremost, the methodologist should be intimately aware of the limitations of the methods under study so that the data-generating process matches the analysis method. Second, there should be extensive \emph{integration testing} of both the data generation function and the analysis function together, for different settings of the simulation factors. Here, it is particularly useful to include a random seed in the generation function so patterns that lead to non-convergence can be investigated. After this process, non-convergence or problematic fit can still remain, and even be deemed acceptable by the methodologist; it can be included as one of the outcomes in the simulation. For example:

\begin{minted}{python}
def analyze_data(df: pl.DataFrame, outcome: str, correction: bool):
    # cast treated column to integer, needed for model fitting
    df = df.with_columns(pl.col.treated.cast(int))
    
    # select columns based on simulation factors
    y = df["post"] - df["pre"] if outcome == "change" else df["post"]
    X = df.select(["treated", "pre"]) if correction else df.select("treated")
    
    # create and fit the model
    mod = sm.OLS(y.to_numpy(), sm.add_constant(X.to_numpy()))
    res = mod.fit()

    # return values of interest, including multicollinearity indicator
    return res.params[1], res.pvalues[1], res.eigenvals[-1] < 1e-10
\end{minted}

The outputs of the analysis function define a single, tidy row of the results table, which is the final component of tidy simulation design, discussed in the following section.

\subsection{Results table}
\label{sec:restab}

Running the data generation (Section \ref{sec:datgen}) and analysis functions (Section \ref{sec:analfun}) for each row of the simulation grid (Section \ref{sec:simgrid}) yields each row of the results table. The results table should again be tidy, meaning each cell represents a single value, and each column represents a single variable of interest (Table \ref{tab:results}). For small simulations, this table can be stored as a data frame in memory, for example like so:

\begin{minted}{python}
# Initialize output list
results_list = []

# iterate over each row in the grid
for row_id, row in enumerate(grid.iter_rows(named=True)):
    df = generate_data(
        sample_size=row["sample_size"],
        effect_size=row["effect_size"],
        seed=row["seed"],
    )
    est, pval, singular = analyze_data(
        df=df, outcome=row["outcome"], correction=row["correction"]
    )
    results_list.append((row_id, est, pval, singular))

# create a tidy results data frame
df_results = pl.DataFrame(
    results_list,
    schema={"row_id": int, "estimate": float, "pvalue": float, "singular": bool},
    orient="row",
)
\end{minted}

For larger, long-running simulations it is necessary to speed up this code. On most computers, this can be done through assigning the work to separate threads or processes. Depending on the hardware of the computer running the simulation, this can speed up the entire simulation dozens of times. In addition, it may be prudent to store output on the hard-disk as the simulation is running, to recover the results in case of an issue with the simulation itself. In the simplest case, this can be done in a text file where results for each row are appended within the main iteration loop. An alternative is to write the results to a dedicated database, which is designed to handle many row-by-row data insertions. This is especially useful if the simulation grid is already part of the same database. See Section \ref{sec:scaling} for more details on scaling tidy simulations.

\begin{table}[ht]
    \centering
    \begin{tabular}{|c|ccc|}
        \hline
        Row ID & Estimate & p-value & Converged \\
        \hline
        1 & 0.01 & 0.973 & true \\
        2 & 0.12 & 0.522 & true \\
        3 & -0.11 & 0.104 & true \\
        $\vdots$ & $\vdots$ & $\vdots$ & $\vdots$ \\
        351998 & 1.03 & 0.001 & true \\
        351999 & 0.99 &  0.019 & true \\
        352000 & 0.96 & 0.002 & true \\
        \hline
    \end{tabular}
    \caption{First and last rows of an example results table. Note that each row from this table corresponds to a row in the simulation grid (Table \ref{tab:sim-grid})}
    \label{tab:results}
\end{table}

Normally, each row in the results table contains just a few values (as in Table \ref{tab:results}). However, there are occasions when the methodologist requires more information than just the output values. An example could be traces of a Markov Chain Monte Carlo (MCMC) algorithm in Bayesian models, which may need to be inspected or processed further. Such additional data can be stored on disk with a unique filename, and the file path can be included as an output in the results table. In this way, post-processing functions can loop over this additional information via the results table (e.g., to compute convergence diagnostics) without taking too much computation during the main simulation and without crowding the results.

As mentioned at the end of section \ref{sec:simgrid}, it some cases it is necessary to perform the simulation in ``wide'' format rather than the standard ``long'' format. To do so means creating more columns in the results table. The methodologist should take care to choose structured column names, which indicate the analysis setting and the outcome value of interest, separated by strings which are easy to parse. For our example simulation, column names could be of the form \texttt{[outcome]\_[correction]\_[value]}, for example \texttt{changescore\_uncorrected\_pvalue}. In the analysis stage (Section \ref{sec:analysis}), this ensures efficient pivoting back to ``long'' format.

\section{Analyzing tidy simulations}
\label{sec:analysis}

The power of tidy simulation shines in the analysis stage. Analyses of tidy simulations can be done after the simulations are done, in an exploratory way after a preliminary number of iterations are run, or even during the simulation in a separate process as the results chunks are chunked output.

The analysis should be a separate script which loads or re-instantiates the simulation grid, and combines it with the results through a simple left-join operation. The most common processing step in the analysis is then to compute summaries of the desired outputs for all conditions aggregated over the Monte Carlo iterations. This can be done with a standard SQL-like \emph{group by} and \emph{aggregate} pattern, as shown below. Note that in the case of a ``wide''-style analysis, the analysis data frame in the following code would first need to be (un)pivoted to a ``long'' format.

\begin{minted}{python}
# load and combine the results
simulation_grid = pl.read_parquet("processed_data/grid.parquet")
results_table = pl.read_parquet("processed_data/results.parquet")
analysis_df = simulation_grid.join(results_table, on="row_id", how="left")

# aggregate to compute power for each condition
df_agg = (
    analysis_df
    .group_by(["sample_size", "effect_size", "outcome", "correction"])
    .agg(
        bias=(pl.col.estimate - pl.col.effect_size).mean(),
        power=(pl.col.pvalue < 0.05).mean(),
    )
)
\end{minted}

It is common to take the mean or a proportion to aggregate over the Monte Carlo simulation error, but for presentation and interpretation it is necessary to also include some form of uncertainty across iterations. For the mean, this could be a quantile (0.025 and 0.975 for lower and upper bound, respectively) or standard error; for a proportion this can be a binomial uncertainty interval based on successes and failures. A detailed, commented example of such uncertainty computation in the aggregation step is shown in Appendix \ref{app:aggregate}.

Since both the analysis data frame and its aggregated version are highly structured due to the tidy nature of the grid and its results, this data frame is naturally compatible with various analysis tools. For example, throught the default formula interface in R, many statistical models are readily applied to this table to statistically analyze the effect of different simulation factors -- just as an actual experiment. An additional feature of this structure is the compatibility with the grammar of graphics \citep{wilkinson2011grammar} as implemented in \texttt{ggplot2} \citep{wickham2011ggplot2} or \texttt{plotnine} \citep{kibirige2025plotnine}. This allows rapid exploratory data analysis and exploration through filtering simulation factor levels, and mapping simulation factors to various aesthetics, for example:

\begin{minted}{python}
import plotnine as p9

plt = (
    p9.ggplot(
        df_agg.filter(pl.col.effect_size.is_in([0.2, 0.5, 0.9])),
        p9.aes(
            x="sample_size",
            y="power",
            ymin="power_lo",
            ymax="power_hi",
            color="outcome",
            linetype="correction",
        ),
    )
    + p9.geom_line()
    + p9.geom_pointrange()
    + p9.facet_wrap("effect_size", labeller="label_both")
    + p9.theme_linedraw()
)
\end{minted}

\begin{figure}[H]
    \centering
    \includegraphics[width=0.9\linewidth]{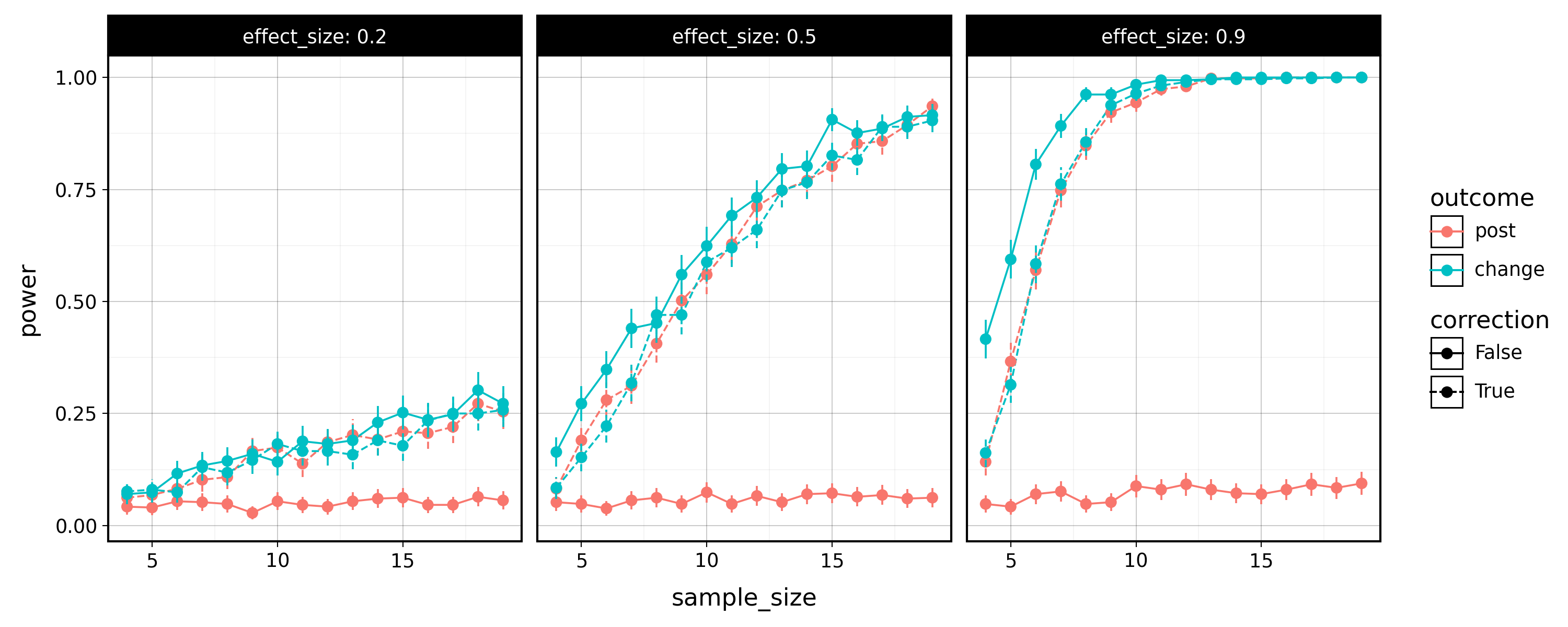}
    \caption{Result of the example simulation, showing that the uncorrected post-measurement outcome analysis does not perform well in terms of power, and the other three methods have similar power for low and medium effect sizes. With a large effect size, the uncorrected change-score analysis has highest power to detect the treatment effect. Note that these conclusions hold only for the specific data-generating process under investigation in the example simulation.}
    \label{fig:result}
\end{figure}

In summary, tidy simulations are compatible and interoperable with common tools for data analysis and visualization in a variety of programming and database languages. This makes these simulations flexible: if a preliminary analysis shows that certain conditions are irrelevant, they can be excluded from the full simulation, which saves time. If, instead, the preliminary analysis shows that more detail is needed in a certain simulation factor, the methodologist can readily add levels to this factor before running the full simulation.

\section{Scaling tidy simulations}
\label{sec:scaling}
Thus far, this paper has presented tidy simulations in the context of a single thread or core on single computer. Several orders of magnitude speedup are possible when this constraint is released. Most modern computers have the possibility to run multiple tasks simultaneously in different processes: 8, 12, 16 or even 32 simultaneous processes are now commonly available on higher-end processors. The tidy simulation framework readily extends to this scalable multi-core mode of operation; large parts of the implementation do not need to be adjusted at all. The fundamental design principle of independence of simulation conditions in Section \ref{sec:simgrid} leads to a so-called ``embarrassingly parallel'' computation: each row represents a task, and the tidy simulation framework does not care which thread, core, or compute node completes this task.

On a single machine, scaling is a matter of assigning rows of the grid as a task to separate parallel processes (e.g., through the parallel package in R or multiprocessing in python) and then collecting the results. An example in python is provided below, and further examples of scaling a tidy simulation in Python and R are available in the online \href{https://github.com/vankesteren/tidy_simulation}{\texttt{code repository}}.

\begin{minted}{python}
from multiprocessing import Pool

# Define a function which returns one row of the results table
def run_simulation(row_id: int, params: dict):
    df = generate_data(
        sample_size=params["sample_size"],
        effect_size=params["effect_size"],
        seed=params["seed"],
    )
    est, pval, singular = analyze_data(
        df=df, outcome=params["outcome"], correction=params["correction"]
    )
    return (row_id, est, pval, singular)


# Run this function in parallel
if __name__ == "__main__":
    with Pool() as p:
        out = p.starmap(
            run_simulation,
            tqdm(enumerate(grid.iter_rows(named=True)), total=len(grid)),
        )
    df = pl.DataFrame(
        out,
        schema={"row_id": int, "estimate": float, "pvalue": float, "singular": bool},
        orient="row",
    )
    df.write_parquet("processed_data/results.parquet")
\end{minted}

The biggest speed-ups can be achieved with large-scale high-performance computing facilities. Scaling a tidy simulation to a compute cluster (or supercomputer) generally requires an additional step of organization beyond single-computer parallel processing: usually, these systems consist of several ``nodes'' which are assigned work efficiently by a scheduler. This can be done using slices (also called chunks, batches) which contain several thousands of simulation grid rows, for example aiming to make each slice on the order of 5 minutes of computation. To run the simulation, the scheduler should assign these chunks of the simulation to different machines, which themselves are often able to do multi-core processing. Compute clusters have a variety of hardware and software available, making it difficult to create generic recommendations; this form of scaling tidy simulations thus falls outside the didactic scope of the present article. Nevertheless, as with scaling to multiple processes on a single computer, the implementation of the grid, the data generation function, and the analysis function do not need to change at all.

\section{Discussion}

This paper introduced tidy simulation: a robust, reproducible, and scalable framework for designing and implementing Monte Carlo simulations. Tidy simulations contain a structured simulation grid representing all required simulation factors, a data generation function implementing the data-generating process under investigation, an analysis function that applies the methods of interest and computes relevant statistics or outcomes. and a structured results table which is the basis of further analysis to provide evidence for the methodological research question. 

While this paper is focused on methodological and statistical research, this structure is not limited to this field: it is quite possible to design a \emph{tidy agent based simulation}. The intermediate functions may change in name or in function, but the general set-up remains, which allows the researcher to implement their simulation as an ``in-silico'' experiment, and analyse its results using common, user-friendly data analysis tools and visualization libraries.

For very computationally intensive methods, the tidy simulation set-up also allows for more advanced extensions such as the aforementioned fractional factorial design of experiments \citep{kelton2000experimental}, surrogate models to estimate intermediate results or expand the input parameter space \citep{barde2024bayesian}, or adaptive sampling with early stopping as is done in some in clinical trials.

Full example code templates of tidy simulation and its analysis are available for R and python at the online \href{https://github.com/vankesteren/tidy_simulation}{\texttt{code repository}}. However, note that this framework is more general than any particular programming language. In general, structured, experimental, and tidy reasoning enable effective and transparent implementation of Monte Carlo simulations. This, in turn, paves the way for high-quality evidence generation for computational statistical science.

\section*{Acknowledgements}
The author thanks \censor{Javier Garcia-Bernardo} for excellent comments on an earlier version of this manuscript, and \censor{Thom Volker} for pointing out relevant additional references.

\section*{Data availability statement}
Data sharing is not applicable to this article as no new data were created or analyzed in this study.

\bibliography{citations}

\begin{thebibliography}{}

\bibitem [\protect \citeauthoryear {%
Alin%
}{%
Alin%
}{%
{\protect \APACyear {2010}}%
}]{%
alin2010multicollinearity}
\APACinsertmetastar {%
alin2010multicollinearity}%
\begin{APACrefauthors}%
Alin, A.%
\end{APACrefauthors}%
\unskip\
\newblock
\APACrefYearMonthDay{2010}{}{}.
\newblock
{\BBOQ}\APACrefatitle {Multicollinearity} {Multicollinearity}.{\BBCQ}
\newblock
\APACjournalVolNumPages{Wiley interdisciplinary reviews: computational statistics}{2}{3}{370--374}.
\PrintBackRefs{\CurrentBib}

\bibitem [\protect \citeauthoryear {%
{Apache Software Foundation}%
}{%
{Apache Software Foundation}%
}{%
{\protect \APACyear {2025}}%
}]{%
apacheparquet}
\APACinsertmetastar {%
apacheparquet}%
\begin{APACrefauthors}%
{Apache Software Foundation}.%
\end{APACrefauthors}%
\unskip\
\newblock
\APACrefYearMonthDay{2025}{}{}.
\newblock
\APACrefbtitle {{P}arquet --- parquet.apache.org.} {{P}arquet --- parquet.apache.org.}
\newblock
\APAChowpublished {\url{https://parquet.apache.org/}}.
\newblock
\APACrefnote{[Accessed 18-07-2025]}
\PrintBackRefs{\CurrentBib}

\bibitem [\protect \citeauthoryear {%
Barde%
}{%
Barde%
}{%
{\protect \APACyear {2024}}%
}]{%
barde2024bayesian}
\APACinsertmetastar {%
barde2024bayesian}%
\begin{APACrefauthors}%
Barde, S.%
\end{APACrefauthors}%
\unskip\
\newblock
\APACrefYearMonthDay{2024}{}{}.
\newblock
{\BBOQ}\APACrefatitle {Bayesian estimation of large-scale simulation models with Gaussian process regression surrogates} {Bayesian estimation of large-scale simulation models with gaussian process regression surrogates}.{\BBCQ}
\newblock
\APACjournalVolNumPages{Computational Statistics \& Data Analysis}{196}{}{107972}.
\PrintBackRefs{\CurrentBib}

\bibitem [\protect \citeauthoryear {%
Besançon%
\ \protect \BOthers {.}}{%
Besançon%
\ \protect \BOthers {.}}{%
{\protect \APACyear {2021}}%
}]{%
besancon2021distributions}
\APACinsertmetastar {%
besancon2021distributions}%
\begin{APACrefauthors}%
Besançon, M.%
, Papamarkou, T.%
, Anthoff, D.%
, Arslan, A.%
, Byrne, S.%
, Lin, D.%
\BCBL {}\ \BBA {} Pearson, J.%
\end{APACrefauthors}%
\unskip\
\newblock
\APACrefYearMonthDay{2021}{}{}.
\newblock
{\BBOQ}\APACrefatitle {Distributions.jl: Definition and Modeling of Probability Distributions in the JuliaStats Ecosystem} {Distributions.jl: Definition and modeling of probability distributions in the juliastats ecosystem}.{\BBCQ}
\newblock
\APACjournalVolNumPages{Journal of Statistical Software}{98}{16}{1--30}.
\newblock
\begin{APACrefURL} \url{https://www.jstatsoft.org/v098/i16} \end{APACrefURL}
\newblock
\begin{APACrefDOI} \doi{10.18637/jss.v098.i16} \end{APACrefDOI}
\PrintBackRefs{\CurrentBib}

\bibitem [\protect \citeauthoryear {%
Boulesteix%
\ \protect \BOthers {.}}{%
Boulesteix%
\ \protect \BOthers {.}}{%
{\protect \APACyear {2020}}%
}]{%
boulesteix2020introduction}
\APACinsertmetastar {%
boulesteix2020introduction}%
\begin{APACrefauthors}%
Boulesteix, A\BHBI L.%
, Groenwold, R\BPBI H.%
, Abrahamowicz, M.%
, Binder, H.%
, Briel, M.%
, Hornung, R.%
\BDBL {}Sauerbrei, W.%
\end{APACrefauthors}%
\unskip\
\newblock
\APACrefYearMonthDay{2020}{}{}.
\newblock
{\BBOQ}\APACrefatitle {Introduction to statistical simulations in health research} {Introduction to statistical simulations in health research}.{\BBCQ}
\newblock
\APACjournalVolNumPages{BMJ open}{10}{12}{e039921}.
\PrintBackRefs{\CurrentBib}

\bibitem [\protect \citeauthoryear {%
Box%
}{%
Box%
}{%
{\protect \APACyear {1976}}%
}]{%
box1976science}
\APACinsertmetastar {%
box1976science}%
\begin{APACrefauthors}%
Box, G\BPBI E.%
\end{APACrefauthors}%
\unskip\
\newblock
\APACrefYearMonthDay{1976}{}{}.
\newblock
{\BBOQ}\APACrefatitle {Science and statistics} {Science and statistics}.{\BBCQ}
\newblock
\APACjournalVolNumPages{Journal of the American Statistical Association}{71}{356}{791--799}.
\PrintBackRefs{\CurrentBib}

\bibitem [\protect \citeauthoryear {%
Brown%
}{%
Brown%
}{%
{\protect \APACyear {2022}}%
}]{%
brown2022simpr}
\APACinsertmetastar {%
brown2022simpr}%
\begin{APACrefauthors}%
Brown, E.%
\end{APACrefauthors}%
\unskip\
\newblock
\APACrefYearMonthDay{2022}{}{}.
\newblock
\APACrefbtitle {simpr: Flexible ’Tidyverse’-Friendly Simulations.} {simpr: Flexible ’tidyverse’-friendly simulations.}
\newblock
\APACaddressPublisher{}{The R Foundation}.
\newblock
\begin{APACrefURL} \url{http://dx.doi.org/10.32614/CRAN.package.simpr} \end{APACrefURL}
\newblock
\begin{APACrefDOI} \doi{10.32614/cran.package.simpr} \end{APACrefDOI}
\PrintBackRefs{\CurrentBib}

\bibitem [\protect \citeauthoryear {%
Burton%
, Altman%
, Royston%
\BCBL {}\ \BBA {} Holder%
}{%
Burton%
\ \protect \BOthers {.}}{%
{\protect \APACyear {2006}}%
}]{%
burton2006design}
\APACinsertmetastar {%
burton2006design}%
\begin{APACrefauthors}%
Burton, A.%
, Altman, D\BPBI G.%
, Royston, P.%
\BCBL {}\ \BBA {} Holder, R\BPBI L.%
\end{APACrefauthors}%
\unskip\
\newblock
\APACrefYearMonthDay{2006}{}{}.
\newblock
{\BBOQ}\APACrefatitle {The design of simulation studies in medical statistics} {The design of simulation studies in medical statistics}.{\BBCQ}
\newblock
\APACjournalVolNumPages{Statistics in medicine}{25}{24}{4279--4292}.
\PrintBackRefs{\CurrentBib}

\bibitem [\protect \citeauthoryear {%
Carriero%
\ \protect \BOthers {.}}{%
Carriero%
\ \protect \BOthers {.}}{%
{\protect \APACyear {2025}}%
}]{%
carriero2025harms}
\APACinsertmetastar {%
carriero2025harms}%
\begin{APACrefauthors}%
Carriero, A.%
, Luijken, K.%
, de Hond, A.%
, Moons, K\BPBI G.%
, van Calster, B.%
\BCBL {}\ \BBA {} van Smeden, M.%
\end{APACrefauthors}%
\unskip\
\newblock
\APACrefYearMonthDay{2025}{}{}.
\newblock
{\BBOQ}\APACrefatitle {The harms of class imbalance corrections for machine learning based prediction models: a simulation study} {The harms of class imbalance corrections for machine learning based prediction models: a simulation study}.{\BBCQ}
\newblock
\APACjournalVolNumPages{Statistics in Medicine}{44}{3-4}{e10320}.
\PrintBackRefs{\CurrentBib}

\bibitem [\protect \citeauthoryear {%
Constantin%
, Schuurman%
\BCBL {}\ \BBA {} Vermunt%
}{%
Constantin%
\ \protect \BOthers {.}}{%
{\protect \APACyear {2023}}%
}]{%
constantin2023general}
\APACinsertmetastar {%
constantin2023general}%
\begin{APACrefauthors}%
Constantin, M\BPBI A.%
, Schuurman, N\BPBI K.%
\BCBL {}\ \BBA {} Vermunt, J\BPBI K.%
\end{APACrefauthors}%
\unskip\
\newblock
\APACrefYearMonthDay{2023}{}{}.
\newblock
{\BBOQ}\APACrefatitle {A general Monte Carlo method for sample size analysis in the context of network models.} {A general monte carlo method for sample size analysis in the context of network models.}{\BBCQ}
\newblock
\APACjournalVolNumPages{Psychological Methods}{}{}{}.
\PrintBackRefs{\CurrentBib}

\bibitem [\protect \citeauthoryear {%
Duncan%
, Tang%
, Elliott%
, Boileau%
\BCBL {}\ \BBA {} Yu%
}{%
Duncan%
\ \protect \BOthers {.}}{%
{\protect \APACyear {2024}}%
}]{%
duncan2024simchef}
\APACinsertmetastar {%
duncan2024simchef}%
\begin{APACrefauthors}%
Duncan, J.%
, Tang, T.%
, Elliott, C\BPBI F.%
, Boileau, P.%
\BCBL {}\ \BBA {} Yu, B.%
\end{APACrefauthors}%
\unskip\
\newblock
\APACrefYearMonthDay{2024}{}{}.
\newblock
{\BBOQ}\APACrefatitle {simChef: High-quality data science simulations in R} {simchef: High-quality data science simulations in r}.{\BBCQ}
\newblock
\APACjournalVolNumPages{Journal of Open Source Software}{9}{95}{6156}.
\newblock
\begin{APACrefURL} \url{https://doi.org/10.21105/joss.06156} \end{APACrefURL}
\newblock
\begin{APACrefDOI} \doi{10.21105/joss.06156} \end{APACrefDOI}
\PrintBackRefs{\CurrentBib}

\bibitem [\protect \citeauthoryear {%
Eager%
\ \BBA {} Roy%
}{%
Eager%
\ \BBA {} Roy%
}{%
{\protect \APACyear {2017}}%
}]{%
eager2017mixed}
\APACinsertmetastar {%
eager2017mixed}%
\begin{APACrefauthors}%
Eager, C.%
\BCBT {}\ \BBA {} Roy, J.%
\end{APACrefauthors}%
\unskip\
\newblock
\APACrefYearMonthDay{2017}{}{}.
\newblock
{\BBOQ}\APACrefatitle {Mixed effects models are sometimes terrible} {Mixed effects models are sometimes terrible}.{\BBCQ}
\newblock
\APACjournalVolNumPages{arXiv preprint arXiv:1701.04858}{}{}{}.
\PrintBackRefs{\CurrentBib}

\bibitem [\protect \citeauthoryear {%
Gelman%
\ \protect \BOthers {.}}{%
Gelman%
\ \protect \BOthers {.}}{%
{\protect \APACyear {2020}}%
}]{%
gelman2020bayesian}
\APACinsertmetastar {%
gelman2020bayesian}%
\begin{APACrefauthors}%
Gelman, A.%
, Vehtari, A.%
, Simpson, D.%
, Margossian, C\BPBI C.%
, Carpenter, B.%
, Yao, Y.%
\BDBL {}Modr{\'a}k, M.%
\end{APACrefauthors}%
\unskip\
\newblock
\APACrefYearMonthDay{2020}{}{}.
\newblock
{\BBOQ}\APACrefatitle {Bayesian workflow} {Bayesian workflow}.{\BBCQ}
\newblock
\APACjournalVolNumPages{arXiv preprint arXiv:2011.01808}{}{}{}.
\PrintBackRefs{\CurrentBib}

\bibitem [\protect \citeauthoryear {%
Gunst%
\ \BBA {} Mason%
}{%
Gunst%
\ \BBA {} Mason%
}{%
{\protect \APACyear {2009}}%
}]{%
gunst2009fractional}
\APACinsertmetastar {%
gunst2009fractional}%
\begin{APACrefauthors}%
Gunst, R\BPBI F.%
\BCBT {}\ \BBA {} Mason, R\BPBI L.%
\end{APACrefauthors}%
\unskip\
\newblock
\APACrefYearMonthDay{2009}{}{}.
\newblock
{\BBOQ}\APACrefatitle {Fractional factorial design} {Fractional factorial design}.{\BBCQ}
\newblock
\APACjournalVolNumPages{Wiley Interdisciplinary Reviews: Computational Statistics}{1}{2}{234--244}.
\PrintBackRefs{\CurrentBib}

\bibitem [\protect \citeauthoryear {%
Harris%
\ \protect \BOthers {.}}{%
Harris%
\ \protect \BOthers {.}}{%
{\protect \APACyear {2020}}%
}]{%
harris2020array}
\APACinsertmetastar {%
harris2020array}%
\begin{APACrefauthors}%
Harris, C\BPBI R.%
, Millman, K\BPBI J.%
, van~der Walt, S\BPBI J.%
, Gommers, R.%
, Virtanen, P.%
, Cournapeau, D.%
\BDBL {}Oliphant, T\BPBI E.%
\end{APACrefauthors}%
\unskip\
\newblock
\APACrefYearMonthDay{2020}{}{}.
\newblock
{\BBOQ}\APACrefatitle {Array programming with {NumPy}} {Array programming with {NumPy}}.{\BBCQ}
\newblock
\APACjournalVolNumPages{Nature}{585}{7825}{357--362}.
\newblock
\begin{APACrefURL} \url{https://doi.org/10.1038/s41586-020-2649-2} \end{APACrefURL}
\newblock
\begin{APACrefDOI} \doi{10.1038/s41586-020-2649-2} \end{APACrefDOI}
\PrintBackRefs{\CurrentBib}

\bibitem [\protect \citeauthoryear {%
Heinze%
}{%
Heinze%
}{%
{\protect \APACyear {2006}}%
}]{%
heinze2006comparative}
\APACinsertmetastar {%
heinze2006comparative}%
\begin{APACrefauthors}%
Heinze, G.%
\end{APACrefauthors}%
\unskip\
\newblock
\APACrefYearMonthDay{2006}{}{}.
\newblock
{\BBOQ}\APACrefatitle {A comparative investigation of methods for logistic regression with separated or nearly separated data} {A comparative investigation of methods for logistic regression with separated or nearly separated data}.{\BBCQ}
\newblock
\APACjournalVolNumPages{Statistics in Medicine}{25}{24}{4216–4226}.
\newblock
\begin{APACrefURL} \url{http://dx.doi.org/10.1002/sim.2687} \end{APACrefURL}
\newblock
\begin{APACrefDOI} \doi{10.1002/sim.2687} \end{APACrefDOI}
\PrintBackRefs{\CurrentBib}

\bibitem [\protect \citeauthoryear {%
Huck%
\ \BBA {} McLean%
}{%
Huck%
\ \BBA {} McLean%
}{%
{\protect \APACyear {1975}}%
}]{%
huck1975using}
\APACinsertmetastar {%
huck1975using}%
\begin{APACrefauthors}%
Huck, S\BPBI W.%
\BCBT {}\ \BBA {} McLean, R\BPBI A.%
\end{APACrefauthors}%
\unskip\
\newblock
\APACrefYearMonthDay{1975}{}{}.
\newblock
{\BBOQ}\APACrefatitle {Using a repeated measures ANOVA to analyze the data from a pretest-posttest design: a potentially confusing task.} {Using a repeated measures anova to analyze the data from a pretest-posttest design: a potentially confusing task.}{\BBCQ}
\newblock
\APACjournalVolNumPages{Psychological bulletin}{82}{4}{511}.
\PrintBackRefs{\CurrentBib}

\bibitem [\protect \citeauthoryear {%
Hussey%
}{%
Hussey%
}{%
{\protect \APACyear {2025}}%
}]{%
hussey2025tidy}
\APACinsertmetastar {%
hussey2025tidy}%
\begin{APACrefauthors}%
Hussey, I.%
\end{APACrefauthors}%
\unskip\
\newblock
\APACrefYearMonthDay{2025}{}{}.
\newblock
\APACrefbtitle {Tidy Monte Carlo simulations.} {Tidy monte carlo simulations.}
\newblock
\APACaddressPublisher{}{Zenodo}.
\newblock
\begin{APACrefURL} \url{https://doi.org/10.5281/zenodo.17224431} \end{APACrefURL}
\newblock
\begin{APACrefDOI} \doi{10.5281/zenodo.17224431} \end{APACrefDOI}
\PrintBackRefs{\CurrentBib}

\bibitem [\protect \citeauthoryear {%
Jamieson%
}{%
Jamieson%
}{%
{\protect \APACyear {1995}}%
}]{%
jamieson1995measurement}
\APACinsertmetastar {%
jamieson1995measurement}%
\begin{APACrefauthors}%
Jamieson, J.%
\end{APACrefauthors}%
\unskip\
\newblock
\APACrefYearMonthDay{1995}{}{}.
\newblock
{\BBOQ}\APACrefatitle {Measurement of change and the law of initial values: A computer simulation study} {Measurement of change and the law of initial values: A computer simulation study}.{\BBCQ}
\newblock
\APACjournalVolNumPages{Educational and Psychological Measurement}{55}{1}{38--46}.
\PrintBackRefs{\CurrentBib}

\bibitem [\protect \citeauthoryear {%
Joshi%
\ \BBA {} Pustejovsky%
}{%
Joshi%
\ \BBA {} Pustejovsky%
}{%
{\protect \APACyear {2025}}%
}]{%
joshi2025simhelpers}
\APACinsertmetastar {%
joshi2025simhelpers}%
\begin{APACrefauthors}%
Joshi, M.%
\BCBT {}\ \BBA {} Pustejovsky, J.%
\end{APACrefauthors}%
\unskip\
\newblock
\APACrefYearMonthDay{2025}{}{}.
\newblock
{\BBOQ}\APACrefatitle {simhelpers: Helper Functions for Simulation Studies} {simhelpers: Helper functions for simulation studies}{\BBCQ}\ [\bibcomputersoftwaremanual].
\newblock
\begin{APACrefURL} \url{https://meghapsimatrix.github.io/simhelpers/} \end{APACrefURL}
\newblock
\APACrefnote{R package version 0.3.1}
\PrintBackRefs{\CurrentBib}

\bibitem [\protect \citeauthoryear {%
Kelton%
}{%
Kelton%
}{%
{\protect \APACyear {2000}}%
}]{%
kelton2000experimental}
\APACinsertmetastar {%
kelton2000experimental}%
\begin{APACrefauthors}%
Kelton, W\BPBI D.%
\end{APACrefauthors}%
\unskip\
\newblock
\APACrefYearMonthDay{2000}{}{}.
\newblock
{\BBOQ}\APACrefatitle {Experimental design for simulation} {Experimental design for simulation}.{\BBCQ}
\newblock
\BIn{} \APACrefbtitle {2000 Winter Simulation Conference Proceedings (Cat. No. 00CH37165)} {2000 winter simulation conference proceedings (cat. no. 00ch37165)}\ (\BVOL~1, \BPGS\ 32--38).
\PrintBackRefs{\CurrentBib}

\bibitem [\protect \citeauthoryear {%
Kisbu-Sakarya%
, MacKinnon%
\BCBL {}\ \BBA {} Aiken%
}{%
Kisbu-Sakarya%
\ \protect \BOthers {.}}{%
{\protect \APACyear {2013}}%
}]{%
kisbu2013monte}
\APACinsertmetastar {%
kisbu2013monte}%
\begin{APACrefauthors}%
Kisbu-Sakarya, Y.%
, MacKinnon, D\BPBI P.%
\BCBL {}\ \BBA {} Aiken, L\BPBI S.%
\end{APACrefauthors}%
\unskip\
\newblock
\APACrefYearMonthDay{2013}{}{}.
\newblock
{\BBOQ}\APACrefatitle {A Monte Carlo comparison study of the power of the analysis of covariance, simple difference, and residual change scores in testing two-wave data} {A monte carlo comparison study of the power of the analysis of covariance, simple difference, and residual change scores in testing two-wave data}.{\BBCQ}
\newblock
\APACjournalVolNumPages{Educational and Psychological Measurement}{73}{1}{47--62}.
\PrintBackRefs{\CurrentBib}

\bibitem [\protect \citeauthoryear {%
Lakens%
\ \BBA {} Caldwell%
}{%
Lakens%
\ \BBA {} Caldwell%
}{%
{\protect \APACyear {2021}}%
}]{%
lakens2021simulation}
\APACinsertmetastar {%
lakens2021simulation}%
\begin{APACrefauthors}%
Lakens, D.%
\BCBT {}\ \BBA {} Caldwell, A\BPBI R.%
\end{APACrefauthors}%
\unskip\
\newblock
\APACrefYearMonthDay{2021}{}{}.
\newblock
{\BBOQ}\APACrefatitle {Simulation-based power analysis for factorial analysis of variance designs} {Simulation-based power analysis for factorial analysis of variance designs}.{\BBCQ}
\newblock
\APACjournalVolNumPages{Advances in Methods and Practices in Psychological Science}{4}{1}{}.
\PrintBackRefs{\CurrentBib}

\bibitem [\protect \citeauthoryear {%
Lee%
, Sriutaisuk%
\BCBL {}\ \BBA {} Kim%
}{%
Lee%
\ \protect \BOthers {.}}{%
{\protect \APACyear {2019}}%
}]{%
lee2019using}
\APACinsertmetastar {%
lee2019using}%
\begin{APACrefauthors}%
Lee, S.%
, Sriutaisuk, S.%
\BCBL {}\ \BBA {} Kim, H.%
\end{APACrefauthors}%
\unskip\
\newblock
\APACrefYearMonthDay{2019}{}{}.
\newblock
{\BBOQ}\APACrefatitle {Using the Tidyverse Package in R for Simulation Studies in SEM} {Using the tidyverse package in r for simulation studies in sem}.{\BBCQ}
\newblock
\APACjournalVolNumPages{Structural Equation Modeling: A Multidisciplinary Journal}{27}{3}{468–482}.
\newblock
\begin{APACrefURL} \url{http://dx.doi.org/10.1080/10705511.2019.1644515} \end{APACrefURL}
\newblock
\begin{APACrefDOI} \doi{10.1080/10705511.2019.1644515} \end{APACrefDOI}
\PrintBackRefs{\CurrentBib}

\bibitem [\protect \citeauthoryear {%
Liu%
, Lu%
, Mogg%
, Mallick%
\BCBL {}\ \BBA {} Mehrotra%
}{%
Liu%
\ \protect \BOthers {.}}{%
{\protect \APACyear {2009}}%
}]{%
liu2009should}
\APACinsertmetastar {%
liu2009should}%
\begin{APACrefauthors}%
Liu, G\BPBI F.%
, Lu, K.%
, Mogg, R.%
, Mallick, M.%
\BCBL {}\ \BBA {} Mehrotra, D\BPBI V.%
\end{APACrefauthors}%
\unskip\
\newblock
\APACrefYearMonthDay{2009}{}{}.
\newblock
{\BBOQ}\APACrefatitle {Should baseline be a covariate or dependent variable in analyses of change from baseline in clinical trials?} {Should baseline be a covariate or dependent variable in analyses of change from baseline in clinical trials?}{\BBCQ}
\newblock
\APACjournalVolNumPages{Statistics in medicine}{28}{20}{2509--2530}.
\PrintBackRefs{\CurrentBib}

\bibitem [\protect \citeauthoryear {%
MacKinnon%
, Warsi%
\BCBL {}\ \BBA {} Dwyer%
}{%
MacKinnon%
\ \protect \BOthers {.}}{%
{\protect \APACyear {1995}}%
}]{%
mackinnon1995simulation}
\APACinsertmetastar {%
mackinnon1995simulation}%
\begin{APACrefauthors}%
MacKinnon, D\BPBI P.%
, Warsi, G.%
\BCBL {}\ \BBA {} Dwyer, J\BPBI H.%
\end{APACrefauthors}%
\unskip\
\newblock
\APACrefYearMonthDay{1995}{}{}.
\newblock
{\BBOQ}\APACrefatitle {A simulation study of mediated effect measures} {A simulation study of mediated effect measures}.{\BBCQ}
\newblock
\APACjournalVolNumPages{Multivariate behavioral research}{30}{1}{41--62}.
\PrintBackRefs{\CurrentBib}

\bibitem [\protect \citeauthoryear {%
Morris%
, White%
\BCBL {}\ \BBA {} Crowther%
}{%
Morris%
\ \protect \BOthers {.}}{%
{\protect \APACyear {2019}}%
}]{%
morris2019using}
\APACinsertmetastar {%
morris2019using}%
\begin{APACrefauthors}%
Morris, T\BPBI P.%
, White, I\BPBI R.%
\BCBL {}\ \BBA {} Crowther, M\BPBI J.%
\end{APACrefauthors}%
\unskip\
\newblock
\APACrefYearMonthDay{2019}{}{}.
\newblock
{\BBOQ}\APACrefatitle {Using simulation studies to evaluate statistical methods} {Using simulation studies to evaluate statistical methods}.{\BBCQ}
\newblock
\APACjournalVolNumPages{Statistics in medicine}{38}{11}{2074--2102}.
\PrintBackRefs{\CurrentBib}

\bibitem [\protect \citeauthoryear {%
Norman%
}{%
Norman%
}{%
{\protect \APACyear {1989}}%
}]{%
norman1989issues}
\APACinsertmetastar {%
norman1989issues}%
\begin{APACrefauthors}%
Norman, G\BPBI R.%
\end{APACrefauthors}%
\unskip\
\newblock
\APACrefYearMonthDay{1989}{}{}.
\newblock
{\BBOQ}\APACrefatitle {Issues in the use of change scores in randomized trials} {Issues in the use of change scores in randomized trials}.{\BBCQ}
\newblock
\APACjournalVolNumPages{Journal of clinical epidemiology}{42}{11}{1097--1105}.
\PrintBackRefs{\CurrentBib}

\bibitem [\protect \citeauthoryear {%
Paxton%
, Curran%
, Bollen%
, Kirby%
\BCBL {}\ \BBA {} Chen%
}{%
Paxton%
\ \protect \BOthers {.}}{%
{\protect \APACyear {2001}}%
}]{%
paxton2001monte}
\APACinsertmetastar {%
paxton2001monte}%
\begin{APACrefauthors}%
Paxton, P.%
, Curran, P\BPBI J.%
, Bollen, K\BPBI A.%
, Kirby, J.%
\BCBL {}\ \BBA {} Chen, F.%
\end{APACrefauthors}%
\unskip\
\newblock
\APACrefYearMonthDay{2001}{}{}.
\newblock
{\BBOQ}\APACrefatitle {Monte Carlo experiments: Design and implementation} {Monte carlo experiments: Design and implementation}.{\BBCQ}
\newblock
\APACjournalVolNumPages{Structural Equation Modeling}{8}{2}{287--312}.
\PrintBackRefs{\CurrentBib}

\bibitem [\protect \citeauthoryear {%
{Plotnine development team}%
}{%
{Plotnine development team}%
}{%
{\protect \APACyear {2025}}%
}]{%
kibirige2025plotnine}
\APACinsertmetastar {%
kibirige2025plotnine}%
\begin{APACrefauthors}%
{Plotnine development team}.%
\end{APACrefauthors}%
\unskip\
\newblock
\APACrefYearMonthDay{2025}{}{}.
\newblock
\APACrefbtitle {plotnine: A grammar of graphics for Python.} {plotnine: A grammar of graphics for python.}
\newblock
\begin{APACrefURL} \url{https://github.com/has2k1/plotnine} \end{APACrefURL}
\newblock
\begin{APACrefDOI} \doi{https://doi.org/10.5281/zenodo.1325308} \end{APACrefDOI}
\PrintBackRefs{\CurrentBib}

\bibitem [\protect \citeauthoryear {%
Raasveldt%
\ \BBA {} M{\"u}hleisen%
}{%
Raasveldt%
\ \BBA {} M{\"u}hleisen%
}{%
{\protect \APACyear {2019}}%
}]{%
raasveldt2019duckdb}
\APACinsertmetastar {%
raasveldt2019duckdb}%
\begin{APACrefauthors}%
Raasveldt, M.%
\BCBT {}\ \BBA {} M{\"u}hleisen, H.%
\end{APACrefauthors}%
\unskip\
\newblock
\APACrefYearMonthDay{2019}{}{}.
\newblock
{\BBOQ}\APACrefatitle {Duckdb: an embeddable analytical database} {Duckdb: an embeddable analytical database}.{\BBCQ}
\newblock
\BIn{} \APACrefbtitle {Proceedings of the 2019 international conference on management of data} {Proceedings of the 2019 international conference on management of data}\ (\BPGS\ 1981--1984).
\PrintBackRefs{\CurrentBib}

\bibitem [\protect \citeauthoryear {%
Tibshirani%
}{%
Tibshirani%
}{%
{\protect \APACyear {1996}}%
}]{%
tibshirani1996regression}
\APACinsertmetastar {%
tibshirani1996regression}%
\begin{APACrefauthors}%
Tibshirani, R.%
\end{APACrefauthors}%
\unskip\
\newblock
\APACrefYearMonthDay{1996}{}{}.
\newblock
{\BBOQ}\APACrefatitle {Regression shrinkage and selection via the lasso} {Regression shrinkage and selection via the lasso}.{\BBCQ}
\newblock
\APACjournalVolNumPages{Journal of the Royal Statistical Society Series B: Statistical Methodology}{58}{1}{267--288}.
\PrintBackRefs{\CurrentBib}

\bibitem [\protect \citeauthoryear {%
Tukey%
}{%
Tukey%
}{%
{\protect \APACyear {1977}}%
}]{%
tukey1977exploratory}
\APACinsertmetastar {%
tukey1977exploratory}%
\begin{APACrefauthors}%
Tukey, J\BPBI W.%
\end{APACrefauthors}%
\unskip\
\newblock
\APACrefYear{1977}.
\newblock
\APACrefbtitle {Exploratory data analysis} {Exploratory data analysis}\ (\BVOL~2).
\newblock
\APACaddressPublisher{}{Springer}.
\PrintBackRefs{\CurrentBib}

\bibitem [\protect \citeauthoryear {%
van Kesteren%
}{%
van Kesteren%
}{%
{\protect \APACyear {2025}}%
}]{%
vankesteren2025polarsgrid}
\APACinsertmetastar {%
vankesteren2025polarsgrid}%
\begin{APACrefauthors}%
van Kesteren, E\BHBI J.%
\end{APACrefauthors}%
\unskip\
\newblock
\APACrefYearMonthDay{2025}{}{}.
\newblock
\APACrefbtitle {vankesteren/polarsgrid: v0.3.0.} {vankesteren/polarsgrid: v0.3.0.}
\newblock
\APACaddressPublisher{}{Zenodo}.
\newblock
\begin{APACrefURL} \url{https://doi.org/10.5281/zenodo.16894273} \end{APACrefURL}
\newblock
\begin{APACrefDOI} \doi{10.5281/zenodo.16894273} \end{APACrefDOI}
\PrintBackRefs{\CurrentBib}

\bibitem [\protect \citeauthoryear {%
Vink%
\ \protect \BOthers {.}}{%
Vink%
\ \protect \BOthers {.}}{%
{\protect \APACyear {2025}}%
}]{%
vink2025polars}
\APACinsertmetastar {%
vink2025polars}%
\begin{APACrefauthors}%
Vink, R.%
, de Gooijer, S.%
, Beedie, A.%
, Burghoorn, G.%
, nameexhaustion%
, Peters, O.%
\BDBL {}Wilksch, M.%
\end{APACrefauthors}%
\unskip\
\newblock
\APACrefYearMonthDay{2025}{}{}.
\newblock
\APACrefbtitle {pola-rs/polars: Python Polars 1.32.2.} {pola-rs/polars: Python polars 1.32.2.}
\newblock
\APACaddressPublisher{}{Zenodo}.
\newblock
\begin{APACrefURL} \url{https://doi.org/10.5281/zenodo.16759930} \end{APACrefURL}
\newblock
\begin{APACrefDOI} \doi{10.5281/zenodo.16759930} \end{APACrefDOI}
\PrintBackRefs{\CurrentBib}

\bibitem [\protect \citeauthoryear {%
Virtanen%
\ \protect \BOthers {.}}{%
Virtanen%
\ \protect \BOthers {.}}{%
{\protect \APACyear {2020}}%
}]{%
virtanen2020scipy}
\APACinsertmetastar {%
virtanen2020scipy}%
\begin{APACrefauthors}%
Virtanen, P.%
, Gommers, R.%
, Oliphant, T\BPBI E.%
, Haberland, M.%
, Reddy, T.%
, Cournapeau, D.%
\BDBL {}{SciPy 1.0 Contributors}%
\end{APACrefauthors}%
\unskip\
\newblock
\APACrefYearMonthDay{2020}{}{}.
\newblock
{\BBOQ}\APACrefatitle {{{SciPy} 1.0: Fundamental Algorithms for Scientific Computing in Python}} {{{SciPy} 1.0: Fundamental Algorithms for Scientific Computing in Python}}.{\BBCQ}
\newblock
\APACjournalVolNumPages{Nature Methods}{17}{}{261--272}.
\newblock
\begin{APACrefDOI} \doi{10.1038/s41592-019-0686-2} \end{APACrefDOI}
\PrintBackRefs{\CurrentBib}

\bibitem [\protect \citeauthoryear {%
Wickham%
}{%
Wickham%
}{%
{\protect \APACyear {2011}}%
}]{%
wickham2011ggplot2}
\APACinsertmetastar {%
wickham2011ggplot2}%
\begin{APACrefauthors}%
Wickham, H.%
\end{APACrefauthors}%
\unskip\
\newblock
\APACrefYearMonthDay{2011}{}{}.
\newblock
{\BBOQ}\APACrefatitle {ggplot2} {ggplot2}.{\BBCQ}
\newblock
\APACjournalVolNumPages{Wiley interdisciplinary reviews: computational statistics}{3}{2}{180--185}.
\PrintBackRefs{\CurrentBib}

\bibitem [\protect \citeauthoryear {%
Wickham%
}{%
Wickham%
}{%
{\protect \APACyear {2014}}%
}]{%
wickham2014tidy}
\APACinsertmetastar {%
wickham2014tidy}%
\begin{APACrefauthors}%
Wickham, H.%
\end{APACrefauthors}%
\unskip\
\newblock
\APACrefYearMonthDay{2014}{}{}.
\newblock
{\BBOQ}\APACrefatitle {Tidy data} {Tidy data}.{\BBCQ}
\newblock
\APACjournalVolNumPages{Journal of statistical software}{59}{}{1--23}.
\PrintBackRefs{\CurrentBib}

\bibitem [\protect \citeauthoryear {%
Wilkinson%
}{%
Wilkinson%
}{%
{\protect \APACyear {2011}}%
}]{%
wilkinson2011grammar}
\APACinsertmetastar {%
wilkinson2011grammar}%
\begin{APACrefauthors}%
Wilkinson, L.%
\end{APACrefauthors}%
\unskip\
\newblock
\APACrefYearMonthDay{2011}{}{}.
\newblock
{\BBOQ}\APACrefatitle {The grammar of graphics} {The grammar of graphics}.{\BBCQ}
\newblock
\BIn{} \APACrefbtitle {Handbook of computational statistics: Concepts and methods} {Handbook of computational statistics: Concepts and methods}\ (\BPGS\ 375--414).
\newblock
\APACaddressPublisher{}{Springer}.
\PrintBackRefs{\CurrentBib}

\bibitem [\protect \citeauthoryear {%
Yoo%
, Jette%
\BCBL {}\ \BBA {} Grondona%
}{%
Yoo%
\ \protect \BOthers {.}}{%
{\protect \APACyear {2003}}%
}]{%
yoo2003slurm}
\APACinsertmetastar {%
yoo2003slurm}%
\begin{APACrefauthors}%
Yoo, A\BPBI B.%
, Jette, M\BPBI A.%
\BCBL {}\ \BBA {} Grondona, M.%
\end{APACrefauthors}%
\unskip\
\newblock
\APACrefYearMonthDay{2003}{}{}.
\newblock
{\BBOQ}\APACrefatitle {Slurm: Simple linux utility for resource management} {Slurm: Simple linux utility for resource management}.{\BBCQ}
\newblock
\BIn{} \APACrefbtitle {Workshop on job scheduling strategies for parallel processing} {Workshop on job scheduling strategies for parallel processing}\ (\BPGS\ 44--60).
\PrintBackRefs{\CurrentBib}

\end{thebibliography}

\appendix

\newpage
\section{Commented aggregation code with uncertainty}
\label{app:aggregate}

The following code aggregates simulation results for both a mean and a proportion and adds uncertainty intervals through computing a quantile and a normal approximation of the binomial confidence interval. It uses the expressive polars dataframe library \citep{vink2025polars}.

\begin{minted}{python}
df_agg = (
    # start with the dataframe of grid parameters and results
    analysis_df
    
    # for each row, compute the bias and whether H0 is rejected
    .with_columns(
        bias=pl.col.estimate - pl.col.effect_size,
        reject=pl.col.pvalue < 0.05
    )
    
    # then group by all the simulation factors
    .group_by(["sample_size", "effect_size", "outcome", "correction"])
    
    # aggregate over iterations, with quantile interval for the bias 
    .agg(
        bias=pl.col.bias.mean(),
        bias_lo=pl.col.bias.quantile(0.025),
        bias_hi=pl.col.bias.quantile(0.975),
        power=pl.col.reject.mean(),
        n_sim=pl.len(),
    )
    
    # then use a normal approximation to compute the CI for power
    .with_columns(
        power_se=(pl.col.power * (1 - pl.col.power) / pl.col.n_sim).sqrt()
    )
    .with_columns(
        power_lo=(pl.col.power - 1.96 * pl.col.power_se).clip(0.0, 1.0),
        power_hi=(pl.col.power + 1.96 * pl.col.power_se).clip(0.0, 1.0),
    )
)
\end{minted}

\end{document}